\documentclass[floatfix,twocolumn,showpacs]{revtex4}
\usepackage{amsmath,amssymb}
\usepackage{amsfonts,amsthm}
\usepackage{epsfig}
\usepackage{graphicx}
\usepackage{dcolumn}
\usepackage{color}
\usepackage{bm}

\begin{document}
\title{Quantum phase transitions in alternating spin-($\frac{1}{2}$, $\frac{5}{2}$) Heisenberg chains}
\author{Ant\^onio S. F. Ten\'orio}
\author{R.~R. Montenegro-Filho}
\author{M.~D. Coutinho-Filho}
\affiliation{Laborat\'orio de F\'{\i}sica Te\'orica e Computacional,
Departamento de F\'{\i}sica, Universidade Federal de Pernambuco, CEP 50670-901,
Recife, Pernambuco, Brazil}

\begin{abstract}
The ground state spin-wave excitations and thermodynamic properties of two types of ferrimagnetic chains
are investigated: the alternating spin-1/2 spin-5/2 chain and a similar chain with a spin-1/2 pendant attached
to the spin-5/2 site. Results for magnetic susceptibility, magnetization and specific heat are obtained through the
finite-temperature Lanczos method with the aim in describing available experimental data, as well as comparison with theoretical
results from the semiclassical approximation and the low-temperature susceptibility expansion derived from Takahashi's modified spin-wave theory. 
In particular, we study in detail the temperature vs. magnetic field phase diagram of the spin-1/2 spin-5/2 chain, 
in which several low-temperature quantum phases are identified: the Luttinger Liquid phase, the 
ferrimagnetic plateau and the fully polarized one, and the respective quantum critical points and crossover lines.
\end{abstract}


\maketitle
\section{Introduction}

Quasi-one-dimensional magnetic materials form a class of compounds with magnetic properties that above a characteristic temperature can be described
through one-dimensional models \cite{giamarchi}. These include systems with a rotationally invariant singlet ground state (GS), modeled, for example, 
through spin-1 gapped and gapless critical spin-1/2 chains \cite{giamarchi}, as well as more complex structures, such us ladders \cite{giamarchi} and spin tubes \cite{IvanovPRL}.
Typically, gapless one-dimensional systems exhibit power-law decay of the correlation functions and can be understood through the 
Bethe ansatz \cite{takalivro} or the Luttinger Liquid theory \cite{giamarchi}. 
In addition, in gapped one-dimensional (1D) systems the application of an external magnetic field $B$ can suppress the gap and induce a quantum phase
transition \cite{QPT} to a Luttinger Liquid phase. In particular, an extensive study of the $B$ - temperature ($T$) phase diagram
of a spin-1/2 gapped ladder system was recently carried out \cite{giaprl2008,giaprb2011}.
 
Contrary to the above mentioned systems, quasi-1D ferrimagnetic compounds display GS spontaneous magnetization and have ferromagnetic 
and antiferromagnetic (AF) spin-wave excitations. 
Usually the AF spin-wave mode is gapped and the magnetization curve exhibits a plateau, which can be explained by topological 
arguments \cite{Oshikawa}. 
Ferrimagnetism can arise from the topology of the unit cell \cite{HalfFilling-doped-Macedo}, as in the phosphates with chemical
 formula A$_3$Cu$_3$(PO$_4$)$_4$,
where A = Ca, Sr or Pb. These materials have three Cu$^{2+}$ spin-1/2 ions \cite{Matsuda} and can be modeled 
by a line of spin-1/2 trimer clusters \cite{Drillon93,Yamamoto2007} with AF exchange couplings.
Another class of ferrimagnets are mixed-spin compounds of type (A-X-B-X-)$_n$, where A
and B are two different magnetic components (single ions or more complex molecules) and X is a bridging ligand. 
In particular, we are interested in compounds that can be modeled by spin-1/2 spin-5/2 chains (sS chains); 
this includes, for example, systems built from Mn$^{2+}$ and Cu$^{2+}$ ions linked through a 
dithioxalato ligand \cite{Gleizes1981, Verdaguer1984}.
Further, in the composition of some ferrimagnets, magnetic elements can be organic radicals like the 
nitronyl nitroxide free radicals (NITR),  where R stands for an alkyl (methyl, ethyl) or
aromatic group (phenyl). A family into this category consists of the Mn-NITR compounds \cite{CaneschiMn}, for which 
there is an AF exchange coupling between the spin-1/2 radicals and the spin-5/2 Mn$^{2+}$ ions. 

In this work we present a numerical study of the thermodynamic properties of the ferrimagnetic chains illustrated in Fig.
\ref{alternatechain}: spin-1/2 spin-5/2 alternating chain (sS chain) and the spin-1/2 spin-5/2 alternating chain with a spin-1/2 pendant
attached to the spin-5/2 site (ssS chain). These chains can be respectively used to model the ferrimagnetic compounds
CuMn(S$_2$C$_2$O$_2$)$_2\text{~$\cdot$~}7.5$H$_2$O (denoted by CuMnDTO) \cite{Verdaguer1984} and
$\mbox{[Mn(NITIm)(NITImH)]ClO}_4$ (denoted by MnNN) \cite{Fegy}, whose crystal structures belong to the 
centrosymmetric monoclinic space group $P2_{1}/c$ ($C_{2h}$). 
The 3D ordered phase observed \cite{Verdaguer1984, Fegy, Verdaguer1987} in these compounds, and in
similar ones \cite{Lhotel}, at very low-$T$ have been intensively investigated. In fact, magnetization measurements \cite{Verdaguer1987,Fegy} suggest
that the canting of the ferrimagnetic moments of the chains give rise to a 3D weak ferromagnetism below the critical temperature, 
although neutron-diffraction experiments \cite{Lhotel} in similar compounds indicate a canted AF structure. A common feature 
in these compounds is that the Mn$^{2+}$ ion has a $^{6}S_{5/2}$ GS, thereby leading to a single ion anisotropy with no zero-field 
splitting in first order of perturbation theory. Very low-$T$ magnetization measurements in a noncentrosymmetric 
orthorhombic compound \cite{Lhotel}, belonging to the space group $P2_{1}2_{1}2_{1}$, 
suggest a single-ion anisotropy $D/k_B\approx 40$ mK, which is much smaller than the intra-chain AF exchange couplings of the 
referred compounds \cite{Verdaguer1984,Verdaguer1987,Fegy,Lhotel}. Therefore, a proper description of the 1D-3D magnetic 
transition may require, in general, anisotropic couplings, including the dipolar interaction. However, similarly to previous
analysis \cite{Verdaguer1984,Fegy}, in order to describe 
the 1D ferrimagnetic properties of the compounds, we disregard anisotropy effects. 

The GS and the low-energy magnetic excitations are calculated through 
the Lanczos exact diagonalization (ED) algorithm, while thermal properties are obtained by the finite-temperature Lanczos method (FTLM) \cite{Jaklic1998}. 
We also explore the field-induced quantum phase transitions of these systems and discuss our results in light of experimental data, as well as predictions from the semiclassical
approximation \cite{Seiden1983} and the modified spin-wave (MSW) theory \cite{takalivro}.

This work will unfold as follows: in Sec. II, we describe the theoretical models and methods employed. In Sec. III, we estimate the model
parameters suitable to describe the experimental data (susceptibility and magnetization) of the related
compounds. In Sec. IV, the one magnon
bands and the specific heats
of the two systems are presented and the main features discussed. In Sec. V we 
exhibit the $T-B$ phase diagram of the sS chain and discuss in detail its
 quantum critical points and crossover lines, 
the Luttinger liquid phase and the plateau regions. 
In Sec. VI, we analyze the low-temperature behavior of the zero-field magnetic susceptibility and, 
finally, in Sec. VII we present a discussion of our relevant findings.

\section{Models and Methods}

\begin{figure}
\begin{center}
\includegraphics*[width=0.30\textwidth,clip]{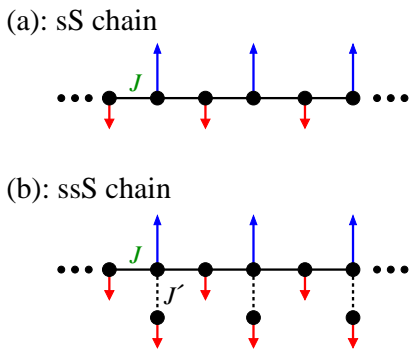}
\caption{(Color online) Schematic representation of the exchange couplings and the GS long-range ferrimagnetic ordering for the (a) sS and (b) ssS alternating chains.}
\label{alternatechain}
\end{center}
\end{figure}
The sS chain with a uniform exchange interaction $J(>0)$ and $N_c$ [s,S] cells is described by the following Hamiltonian:
\begin{equation}
H_{sS} = J\sum_{i}^{N_c}(\mathbf{S}_i + \mathbf{S}_{i+1})\cdot \mathbf{s}_i-g\mu_BB S^z,
\label{HsS}
\end{equation}
where $s=1/2$ and $S=5/2$, the $g$-factor is assumed uniform, $\mu_B$ is the Bohr magneton, $B$ is an applied magnetic
field in the $z$ direction and $S^z$ is the operator for the $z$ component of the total spin. This chain is bipartite, with $N_c$ sites with spin-5/2 in one
sublattice and $N_c$ sites with spin-1/2 in the other; therefore, the Lieb and Mattis theorem \cite{Lieb} assures that the GS total spin, $S_{GS}$,
is given by $N_c|S-s|=2N_c$, i. e., spin 2 per unit cell. The GS magnetic ordering of this chain is sketched in Fig. \ref{alternatechain}(a).

The ssS chain with $N_c$ [s,S,s$^\prime$] cells is described by the following Hamiltonian:
\begin{equation}
H_{ssS}=J\sum_{l=1}^{N_c}\mathbf{s}_l\cdot(\mathbf{S}_{l}+\mathbf{S}_{l+1})+J^{'}\sum_{l=1}^{N_c}
\mathbf{S}_l\cdot\mathbf{s}^{'}_{l}-g\mu_B B S^z,
\label{HssS}
\end{equation}
where $s=s^{\prime}=1/2$ and $S=5/2$, while $J>0$ and $J^{\prime}>0$. The Lieb and Mattis theorem assures
that $S_{GS}=N_c|S-2s|=3N_c/2$, i. e., spin $\frac{3}{2}$ per unit cell. The GS magnetic order of this chain
is sketched in Fig. \ref{alternatechain}(b).

The FTLM \cite{Jaklic1998} is based on the Lanczos diagonalization technique and random sampling.
The fundamental relations used in FTLM for the calculation of an static quantity associated
 to an operator $A$ are
\begin{eqnarray}
\langle A \rangle &\approx&
\frac{N_{st}}{ZR}\sum_{r=1}^{R}\sum_{j=0}^\mathcal{M}
e^{-\beta \epsilon^r_j}\langle r|\psi^r_j\rangle\langle\psi^r_j|A|
r \rangle, \nonumber \\
Z &\approx& \frac{N_{st}}{R}\sum_{r=1}^{R}\sum_{j=0}^\mathcal{M} e^{-\beta
\epsilon^r_j}|\langle r|\psi^r_j\rangle|^2,
\label{L14}
\end{eqnarray}
where the sampling is carried over $R$ random states $|r\rangle$, taken as initial states for
a $\mathcal{M}$-step Lanczos procedure which results in $\mathcal{M}$ approximate eigenvalues $\epsilon^r_j$ with respective
eigenvectors $|\psi^r_j\rangle$ in the $N_{st}$-dimensional Hilbert space.
The method allow us to calculate the temperature dependence of the magnetization per unit cell $m_c$,
magnetic susceptibility per unit cell $\chi$, and specific heat $C$ through:
$
m_c = g\mu_B \frac{\langle S^z\rangle}{N_c},
$
$
\chi = g^2\mu_B^2 \frac{\langle (S^{z})^2\rangle - \langle S^z \rangle^2}{N_ck_BT},\text{ and}
$
$
C = \frac{\langle H^2\rangle - \langle H \rangle^2}{k_BT^2},
$
where $k_B$ is the Boltzmann constant. The total number of sites $N=2N_c$ for the sS chain 
and $N=3N_c$ for the ssS chain.
In the computation we have used periodic boundary conditions, $\mathcal{M}=50$
for both chains and $R=40000$ (50000) for the sS (ssS) chain. 
A full diagonalization study of the specific heat and susceptibility for 
the sS chain with $N_c=3$ can be found in Ref. \cite{Drillon89}. 

\section{Magnetic Susceptibility and Model Parameters}
\begin{figure}
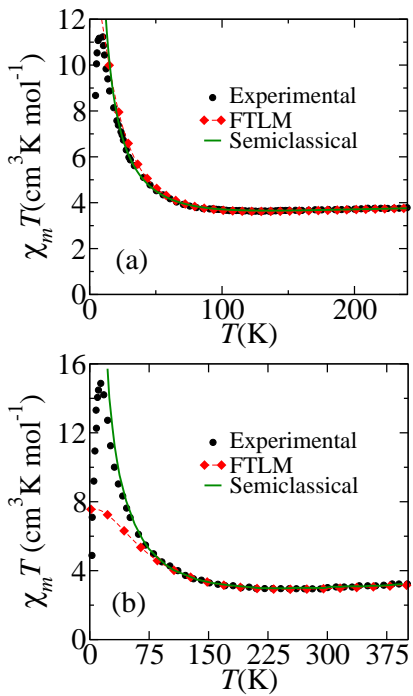

\begin{center}
\includegraphics[width=0.3\textwidth,clip]{fig2a}
\includegraphics[width=0.3\textwidth,clip]{fig2b}
\caption{(Color online) Molar susceptibility $\chi_m$ times temperature $T$ of the
(a) sS and (b) ssS chains as a function of $T$. 
Experimental: data of (a) the compound CuMnDTO from Ref. \cite{Verdaguer1984} and of (b) the MnNN
compound from Ref. \cite{Fegy}.
FTLM: (a) $N=16$, $g=1.90$ and $J=44.8K$, (b) $N=18$, $g=2.0$, $J/k_B=150$ K and $J^{\prime}/k_B=255$ K.
Semiclassical: (a) Ref. \cite{Seiden1983} (with $J/k_B=$ 59.7 K, $S=2.5$ and $g=1.9$) and (b)
Ref. \cite{Fegy} (with $J/k_B=$ 141 K, $J^{'}/k_B=$ 250 K, $S=2.5$ and $g=2.0$).}
\label{chit}
\end{center}
\end{figure}
Through a semiclassical approach, in which the S spins are treated as classical variables, Seiden \cite{Seiden1983}
derived a closed formula for the magnetic susceptibility $\chi$. In particular, the quantity $T\chi(T)$ has a minimum at a
temperature $T_{min}$ which is generally situated
in a region where $\beta JS<1$, a feature which has been known to be typical of 1D ferrimagnets.
Similarly, a closed expression for the susceptibility of the ssS chain can also be established \cite{Fegy}.

In Fig. \ref{chit}(a) we present data for the magnetic susceptibility of the compound CuMnDTO (from Refs. \cite{Verdaguer1984,Seiden1983})
together with FTLM, with $J/k_B=44.8$ K and $g=1.90$, and semiclassical-approximation \cite{Seiden1983} results, with $J/k_B=59.7$ K and $g=1.9$,
for the sS chain. For the FTLM results, the estimation of $J$ is made by using the value
of the minimum of the experimental curve: $T_{min} = 130$ K. We see that both the FTLM and the semiclassical approach agree
with the experimental data in the mid- and high-temperature regimes.
As the temperature is lowered below $T_{min}$, $\chi T$ increases and presents a maximum at $T_{max}=7.5$ K, which marks the
onset of the tridimensional ordering. For a strictly one-dimensional system it is expected, from the Mermin-Wagner theorem \cite{Mermin}, 
that long-range order (LRO) may occur only at $T=0$. We remark that for quantum ferromagnetic \cite{Takahashi1987} chains  
the correlation length diverges as $1/T$ and the susceptibility as $1/T^2$; further, the low-lying magnetic excitations 
of ferrimagnetic chains present a ferromagnetic character (see below) and the same referred critical behavior is shown \cite{EPRMDCF} 
to hold, which explains the increase in the curve of $\chi T$ just below $T=T_{min}$. 
 
In Fig. \ref{chit}(b) we present FTLM and semiclassical results \cite{Fegy} for the ssS
 chain, and experimental data \cite{Fegy} of the MnNN compound. A profile similar to that of Fig. \ref{chit}(a) is observed with $T_{min}=255$ K.
Taking $g=2.0$, our estimative for the model parameters, $J$ and $J^{\prime}$, are $J^{\prime}=1.7J$, with $J/k_B=150$ K and
 $J^{\prime}/k_B=255$ K. We also note that the FTLM curve for this chain depart from the experimental one at a higher temperature
than the curve for the sS chain. This behavior is in fact a finite size effect since the number of unit cells used in the FTLM calculation
for the ssS chain (6 unit cells, with 18 sites) is effectively less than the number used for the sS chain (8 unit cells, with 16 sites).

\begin{figure}
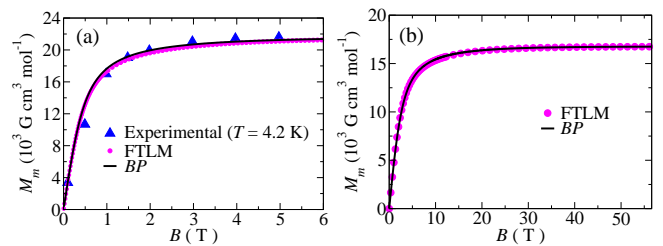

\begin{center}
\includegraphics[width=0.235\textwidth,clip]{fig3a}
\includegraphics[width=0.235\textwidth,clip]{fig3b}
\caption{(Color online) (a) Molar magnetization $M_m$ of the sS chain as a function of $B$ at $T=4.2$ K. 
Experimental data of the compound CuMnDTO from Ref. \cite{Verdaguer1984}. FTLM
 results for a chain with $N=16$, $J/k_B=44.8$ K and $g=1.93$.  Brillouin paramagnet (\textit{BP}) with total spin $\overline{S}=15.0$.
(b) FTLM results for $M_m$ of the ssS chain with $N=18$ as a function of $B$ at $T=15$ K. 
\textit{BP} with $\overline{S}=8.5$.}
\label{mh1}
\end{center}
\end{figure}

In Fig. \ref{mh1} we compare our results at $T=4.2$ K with experimental data from Ref. \cite{Verdaguer1984}.
The temperature is lower than the one in which the maximum of the $\chi T$ curve is observed, $T_{max}\approx 7.5$ K. 
However, due to the low value of
the coupling between chains $J_{\text{inter-chain}}/k_B\sim 0.1$ K, we expect that at $T=4.2$ K and for fields higher than $\sim 0.1$ T, 
the ferrimagnetic correlations along the chain 
are the relevant ones to determine the behavior of the
magnetization as a function of $B$. Since the correlation length along the chains  diverges \cite{EPRMDCF,vitoriano2002} as $1/T$, 
a finite number of unit cells are correlated at 4.2 K. Thus, we can treat the system as composed
 of independent linear clusters, 
each cluster carrying a total spin $\overline{S}$, and a superparamagnetic behavior is expected for the magnetization curve. 
Within this context, we try to estimate the number of correlated cells in the chain from the experimental data shown in
Fig. \ref{mh1}(b) by comparing it with the FTLM data and the molar magnetization of a Brillouin paramagnet (BP) with total spin $\overline{S}$, given by
$M_m(B,T)=  N_Ag\mu_B(S-s)B_{\overline{S}}(x)$,
where $N_A$ is the Avogadro constant and $B_{\overline{S}}(x)$ is the Brillouin function.
As shown in Fig. \ref{mh1}(b), the experimental data is well described by the BP curve with $\overline{S}=15$,
indicating that approximately 8 unit cells (size used in the FTLM calculation) are ferrimagnetically correlated at
4.2 K. This 
enforces the one-dimensional description
of the experimental magnetization for this temperature and field values, as well as the
 superparamagnetic behavior. We remark that the
 authors of Ref. \cite{Verdaguer1984} estimate that approximately 10 cells 
of the compound CuMnDTO are ferrimagnetically correlated at $T=7.9$ K (just above $T_{max})$ for $B=0$.

For the ssS system, we find no published experimental
data for the magnetization. However, considering the FTLM results, Fig. \ref{mh1}(b), we
estimate that at $T=15$ K the number of ferrimagnetically correlated unit cells
is $\sim 6$ (size used in the FTLM calculation), due to the good agreement
between the FTLM results and the BP curve with $\overline{S}=8.75$.

\section{One-magnon Bands and Specific Heat}

\begin{figure}
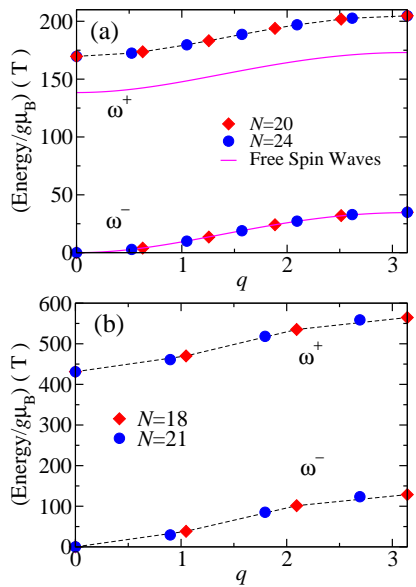

\includegraphics[width=0.3\textwidth,clip]{fig4a}
\includegraphics[width=0.3\textwidth,clip]{fig4b}
\caption{(Color online) ED results for the lower energy one-magnon bands, in units of magnetic field (using $g=1.93$), of the (a) sS and (b) ssS chains for
the indicated values of $N$. Full lines are non-interacting spin-wave results from Ref. \cite{Pati1997}, while dashed lines
 are guide to the eyes.}
\label{magnons}
\end{figure}
Due to the ferrimagnetic order of the GS, there are two kinds of elementary excitations in the systems: ferromagnetic magnons, which
lowers the total spin by one unit and AF magnons, which increases the total spin by one unit. The dispersion relations of the
lower energy magnons are calculated, respectively, through
\begin{eqnarray}
\omega^{-}(q)&=&E_{min}(S_{GS}-1,q)-E_{GS}\\
\omega^{+}(q)&=&E_{min}(S_{GS}+1,q)-E_{GS},
\end{eqnarray}
where $E_{min}(S_t,q)$ indicates the lowest energy in the total-spin sector $S_t$ and lattice wavenumber $q=2\pi l/N_c,\text{ with } l=0,1,2,...,N_c-1$.

In Figs. \ref{magnons}(a) and \ref{magnons}(b) we display the lower energy one-magnon bands, in units of magnetic field, of the sS and ssS chains.
For the sS chain, we also plot in Fig. \ref{magnons}(a) non-interacting spin-wave (SW) results \cite{Pati1997}:
\begin{eqnarray}
\omega^{-}_{SW}(q) & = & -J(S-s)+\omega_q+g\mu_B B, \label{dispersionF}\\
\omega^{+}_{SW}(q) & = & J(S-s)+\omega_q-g\mu_B B, \label{dispersion}
\end{eqnarray}
where $\omega_{q}  =  J\sqrt{(S-s)^2+4sS\sin^2(q/2)}$, $s=1/2$ and
$S=5/2$. We notice that in zero field the ferromagnetic excitation is gapless
(which is expected from the spontaneously broken symmetry of the
GS) and display a quadratic dispersion relation in the long wavelength 
limit, as predicted by conformal invariance \cite{Malvezzi}, while 
a gap $\Delta$ exists for the AF excitation. The ferromagnetic branch obtained
through non-interacting SW theory for the sS chain is in good agreement with the
ED data, while for the AF branch the value of 
the zero field gap $\Delta=2J(S-s)=4J$ departures from the ED value,
as is often the case in other ferrimagnetic systems \cite{otherferri}, due to quantum fluctuations effects. In fact,
 we estimate that in
the thermodynamic limit $\Delta=4.9046J\text{ }(3.88J)$ for the sS
(ssS) chain. 

\begin{figure}
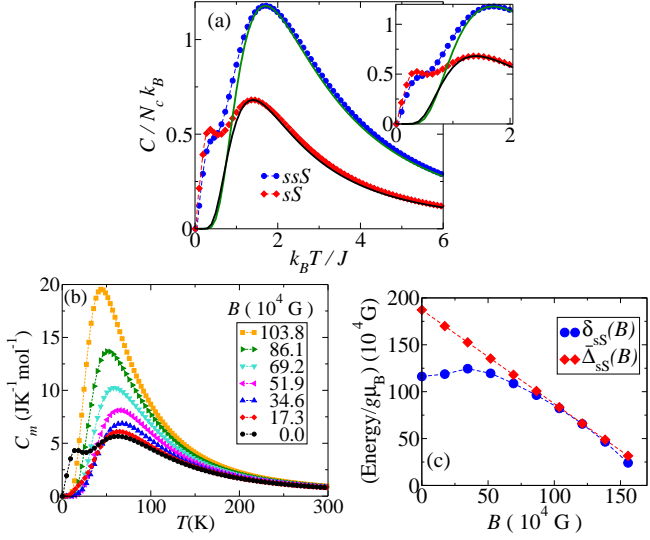

\includegraphics[width=0.28\textwidth,clip]{fig5a}
\includegraphics[width=0.25\textwidth,clip]{fig5b}
\includegraphics[width=0.22\textwidth,clip]{fig5c}
\caption{(Color online) FTLM results. (a) Specific heat of the sS and ssS chains for $B=0$ as a function of temperature $T$;
full lines are the respective Schottky specific heats. (b) Molar specific heat $C_m$ of the sS chain for the indicated
 values of $B$ (using $g=1.93$); (c)
$B$-dependent Schottky gap $\delta_{sS}(B)$ and spin-wave gap $\bar{\Delta}_{sS}(B)$. Dashed lines are guide to the eyes.}
\label{C}
\end{figure}

In Fig. \ref{C}(a) we show the specific heat of the sS and ssS chains in zero field. 
Due to the LRO ferrimagnetic state at $T=0=B$, with low-energy 
gapless ferromagnetic excitations, it is expected that $C\sim \sqrt{T}$.
Another feature is the occurrence of double peaks \cite{Yamamoto1998R, Nakanishi, polonica};
it turns out that the main peak is well described by the Schottky
formula \cite{Yamamoto1998R, Nakanishi}:
$\frac{C}{N_ck_B}=\frac{A(\delta/2k_BT)^2}{\cosh^2(\delta/2k_BT)}$, where $\delta$ is the Schottky gap and $A$ is the amplitude parameter. 
The Schottky gap for the ssS chain
$\delta_{ssS} \approx 4.1J$ is in accord with the AF spin-wave gap $\Delta_{ssS}\approx 3.9J$. However, for the sS chain
the value of $\delta_{sS}$ ($\approx 3.4J$) significantly departures from the AF spin-wave gap value: $\Delta_{sS}\approx4.9J$, indicating 
strong influence of the lower-energy ferromagnetic excitations.
Since these states have a total spin ($S_t=S_g-1$) lower than the one of the AF branch ($S_t=S_g+1$),
we expect that a field $B$ can wash it out. In fact, the center of the AF branch
[see Fig. \ref{magnons}(a)] is found at $\bar{\Delta}_{sS}(0)=5.4J$ and is lowered in the
presence of a magnetic field through $\bar{\Delta}_{sS}(B) = \bar{\Delta}_{sS}(0) - g\mu_B B$. In Fig. \ref{C}(b) we present the
specific heat for fields up to 103.8 T, and in Fig. \ref{C}(c) we compare
$\bar{\Delta}_{sS}(B)$ with the Schottky gap $\delta_{sS}(B)$. As we can see in the figure, the values of the two
quantities are nearly equal for moderate values of $B$. 

\section{T - B Phase Diagram}

The GS magnetization per unit cell, $m_c$, of
one-dimensional systems under an applied magnetic field can exhibit plateaus 
at values such that $S_c-m_c=\text{integer}$, where $S_c$
 is the maximum total spin of a unit cell \cite{Oshikawa}.
This condition implies that a plateau can be observed at values of $m_c$ differing from its saturated 
value by an integer number of spin flips. In particular, a magnetization plateau at 1/3 of the saturation magnetization was observed
in the magnetization curve of the mineral azurite \cite{azurite}, which is generally modeled through
the distorted diamond chain \cite{DistChain}. 
Other compounds exhibiting the 1/3 magnetization 
plateau are the trimer chain systems Cu$_2$(P$_2$O$_6$OH)$_2$ \cite{Hase} and the phosphates \cite{Matsuda}  
A$_3$Cu$_3$(PO$_4$)$_4$, where A = Ca, Sr or Pb. Further, the thermal properties of a variety of 
models \cite{suprb2010,suprb2009,suprb2006} presenting plateaus in their magnetization curves were analyzed in recent 
years and it was evidenced that the 1/3 magnetization plateau is also a characteristic feature of
frustrated spin-S chains \cite{dagottofrust}.

\begin{figure}
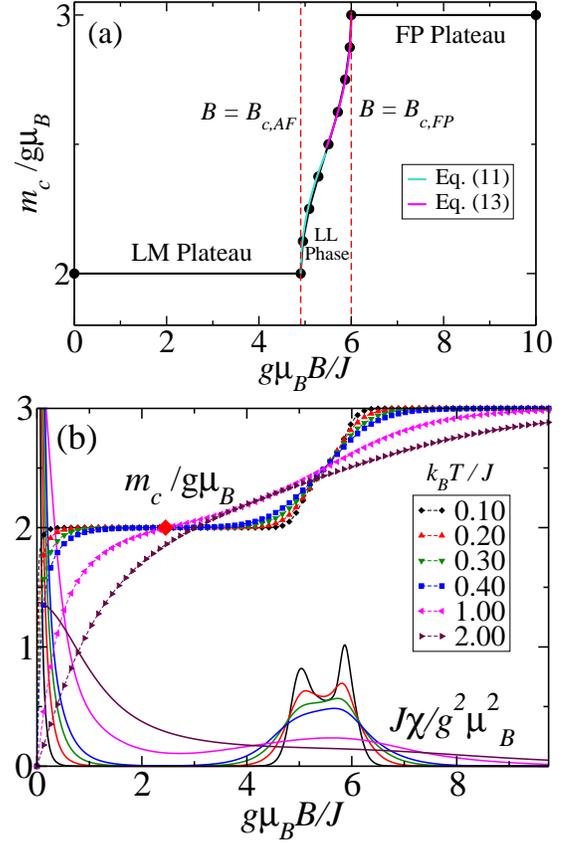

\begin{center}
\includegraphics*[width=0.4\textwidth,clip]{fig6a.eps}
\includegraphics*[width=0.4\textwidth,clip]{fig6b.eps}
\caption{(Color online) FTLM results for the sS chain. (a) GS magnetization per cell $m_c$ as a function of $B$ for $N=16$: 
full circles indicate the midpoints in the steps of the magnetization of the finite-size system and 
edges of thermodynamic-limit plateaus \cite{bonner}, while colored full lines indicate the results from the 
free-fermion model (see text). (b) $m_c$ (curves with symbols) and
$\chi$ (full curves) as a function of $B$ for the listed values of $T$; $B_m$ is 
indicated by the red diamond. }
\label{msS}
\end{center}
\end{figure}
For the sS chain studied here, possible plateaus should be observed at $m_c=m_{LM}=g\mu_B(S-s)$ [Lieb-Mattis (LM) magnetization] and 
$m_c=m_{FP}=g\mu_B(S+s)$ (fully polarized magnetization), as confirmed by the numerical results shown in Fig. \ref{msS} (a).
In zero field the GS is ferrimagnetic with gapless ferromagnetic excitations [Eq. (\ref{dispersionF})], while
for $B\neq0$ this mode acquires     
the gap $\Delta_F(B)=g\mu_B B$. Also, as $B$ increases, the gap for the 
AF mode [Eq. (\ref{dispersion})] decreases linearly with $B$: $\Delta_{AF}(B)=\Delta_{sS}-g\mu_B B$, and 
for $g\mu_B B\geq g\mu_B B_m = (\Delta_{sS}/2)\approx 2.45J$ its gap is equal to the ferromagnetic one. 
At $g\mu_B B=\Delta_{sS}\equiv g\mu_B B_{c,AF}$ the AF gap vanishes and the system undergoes a quantum phase transition
(\textit{condensation of AF magnons}, each carrying a spin +1) to a gapless Luttinger Liquid (LL) phase \cite{ChitraGiamarchi},
 with power-law 
decay of the transverse correlation functions. In fact, the 
quantum critical point $B_{c,AF}$ separates an incompressible phase (plateau) from a compressible one (LL phase). 
For $B\gtrsim B_{c,AF}$, a low-density of magnons is found in the system and the asymptotic singular form of the magnetization can
be obtained \cite{Affleck91} by considering the system as a free Fermi gas or hard-core bosons. 
In this limit, the magnons will occupy single particle states with $q\rightarrow 0$ and the dispersion relation, 
Eq. (\ref{dispersion}), can be used by replacing the linear spin-wave gap, $\Delta_{SW}$, by the computed gap in
Fig. \ref{magnons}(a), $\Delta_{sS}=4.9046J$:
\begin{equation}
\omega_{AF}^{+}=-\mu+\frac{v^2}{2\Delta_{sS}}q^2,\text{ $q\rightarrow 0$,}
\label{q0}
\end{equation}
where $v=J\sqrt{2sS}=J\sqrt{5/2}$, $\mu=g\mu_B B-\Delta_{sS}=g\mu_B(B-B_{c,AF})$ and $\Delta_{SW}=2J(S-s)$. 
The energy density can thus be written (fermionic map) as
\begin{equation}
\varepsilon=\int^{k_F}_{-k_F}\frac{dk}{2\pi}(\epsilon_k-\mu),
\end{equation}
where $\epsilon_k=v^2k^2/2\Delta_{sS}$, $k_F=\pi n$, and $n$ is the density of particles. The value of $n$ for a prescribed $\mu$ can 
be obtained from the condition $\partial_n\varepsilon=0$:
\begin{equation}
n=\sqrt{g\mu_B}\sqrt{\frac{2 B_{c,AF}}{\pi^2 v^2}}\sqrt{\mu},
\label{sing}
\end{equation}
which implies
\begin{equation}
\frac{m_c}{g\mu_B}=2+\frac{g\mu_B}{J}\sqrt{\frac{4B_{c,AF}}{5\pi^2}}\sqrt{B-B_{c,AF}}.
\label{maf}
\end{equation}
In Fig. \ref{msS}(a) we show the very good agreement between the numerical data and $m_c$
given by Eq. (\ref{maf}) in the LL Phase.  
\begin{figure}
\begin{center}
\includegraphics*[width=0.45\textwidth,clip]{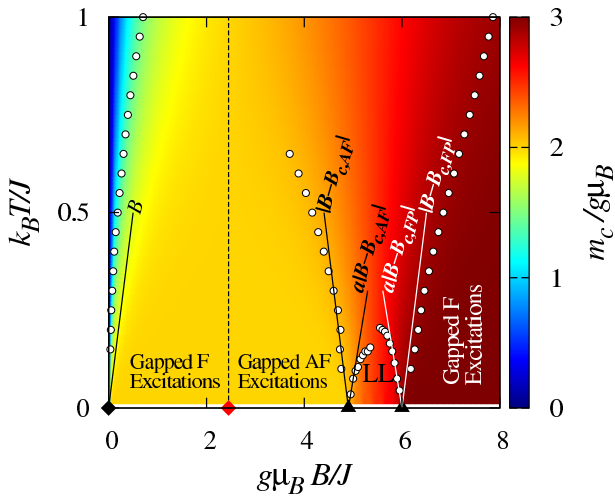}
\caption{(Color online) FTLM results for the low-$T$ phase diagram of the sS chain: contour plot indicates the magnetization per cell $m_c$. The critical
point at $B=0$ (black diamond), the inflection point of $m_c$ at $g\mu_B B=g\mu_B B_m=\Delta/2$ (red diamond), the quantum-critical points (black triangles) 
at $B=B_{c,AF}$ and $B=B_{c,FP}$, crossover lines (white circles) and their asymptotic behavior (full lines) are also indicated.}
\label{DiagM}
\end{center}
\end{figure}

The gapless LL phase ends at the quantum critical point $B=B_{c,FP}$: the system becomes
 fully polarized (FP) and 
presents gapped low energy excitations. The two one-magnon excitations from the FP state, both carrying a spin -1, 
can be exactly obtained \cite{YFerro} and the lower one has a dispersion relation given by
\begin{equation}
\omega_{FP}=-J(s+S)-J\sqrt{(S-s)^2+4sS\cos^2(q/2)}+g\mu_B B,
\label{fp}
\end{equation}
which implies $g\mu_B B_{c,FP}=2J(s+S)=6J$, in accord with the numerical results [Fig. \ref{msS}(a)]. 
For $B\lesssim B_{c,FP}$, a low density of magnons is observed in the system and the same arguments used to obtain 
Eq. (\ref{sing}) can be used in this case. For $q\rightarrow 0$, Eq. (\ref{fp}) can be written as Eq. (\ref{q0}) with 
$v=J\sqrt{2sS}=J\sqrt{5/2}$, $\Delta_{F}=2J(s+S)=6J$ and $\mu=g\mu_B(B_{c,FP}-B)$, which implies, from Eq. (\ref{sing}), 
that the density of magnons is given by $n=\frac{g\mu_B}{J}\sqrt{\frac{4B_{c,FP}}{5\pi^2}}\sqrt{B_{c,FP}-B}$, and $m_c$ now reads: 
\begin{equation}
\frac{m_c}{g\mu_B}=3-\frac{g\mu_B}{J}\sqrt{\frac{4B_{c,FP}}{5\pi^2}}\sqrt{B_{c,FP}-B}, 
\label{ms}
\end{equation}
which is plotted in Fig. \ref{msS}(a) and is also in very good agreement with the numerical data. 

In Fig. \ref{msS}(b) we present FTLM data for $m_c$ and $\chi$ vs. $B$ for $T\neq0$. We first notice that the magnetization  
in zero field is null and the system is in the thermal paramagnetic state, as expected from the Mermin-Wagner theorem \cite{Mermin}.
 Increasing $B$ in the low-temperature regime, 
the LM (or ferrimagnetic) plateau is exponentially 
reached and the magnetization exhibits an inflection point at $B=B_m$ [red diamond in Fig. \ref{msS}(b)] that marks the changing of the gapped low-energy excitations from 
ferromagnetic ($B\leq B_m$) to AF magnons ($B\geq B_m$): the ferromagnetic (antiferromagnetic) magnons are exponentially activated and $m_c$
is lower (higher) than $m_{LM}=g\mu_B(S-s)$. Also, by the same token, the FP plateau is exponentially reached from below
for fields higher than $B_{c,FP}$. Furthermore, the singular form of the magnetization near the quantum critical points ($B=B_{c,AF}$ and $B=B_{c,FP}$, with $T=0$),
which implies $\chi\rightarrow\infty$, are thermally smoothed out and the singularities in the susceptibility evolve into local maxima, 
thus providing the determination of the {\it crossover} lines.
The LL phase, with linear dispersion relation $\sim q$, is expected \cite{ChitraGiamarchi} between the two local maxima for a given $T$ 
[see, e. g., the susceptibility curves for
 $k_B T=0.10J$ and $0.20J$ in Fig. \ref{msS}(b)] with the two local maxima indicating a crossover to a region in which the
 excitations follow a non-relativistic 
dispersion relation $\sim q^2$, as previously discussed. On the other hand, as $T$ increases, the LL 
phase ends and a single maximum is observed in the susceptibility
curves (see, e. g., the susceptibility for $k_B T=0.40J$). This single maximum  defines a crossover from 
the regime in which the physics is determined by the excitations from the LM plateau to a regime in which the FP plateau is 
the relevant one. For sufficiently high temperatures, the system looses all
information about the $T=0$ LM magnetization plateau and the effect of $B$ is to bring the system from
 the thermal paramagnetic state to the FP state at higher magnetic fields (see the case $k_B T=2.00J$).
\begin{figure}
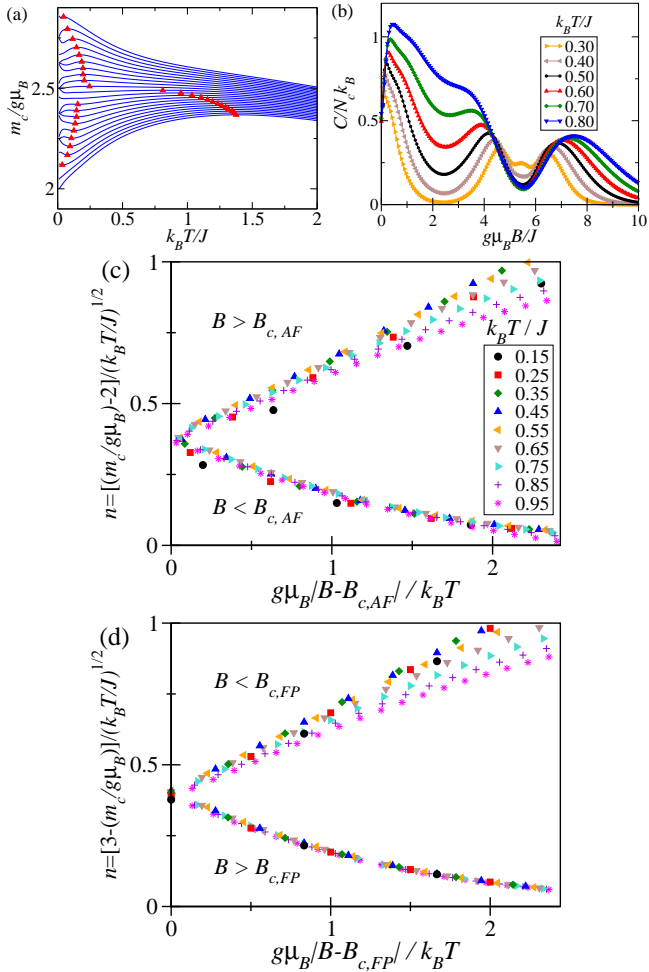

\begin{center}
\includegraphics[width=0.235\textwidth,clip]{fig8a.eps}
\includegraphics[width=0.235\textwidth,clip]{fig8b.eps}
\includegraphics[width=0.35\textwidth,clip]{fig8c.eps}
\includegraphics[width=0.35\textwidth,clip]{fig8d.eps}
\caption{(Color online) FTLM results for the sS chain. (a) Magnetization per cell $m_c$ as a function of $T$ for fields 
$g\mu_B B/J$ from $4.85$ to $5.95$ in steps of $0.05$ (from below to top); triangles indicate local maxima associated with the LL crossover lines (see text).
(b) Specific heat per cell $C$ as a function of $B$ for the indicated values of temperature. 
Scaling of the magnon density $n$ around the quantum critical points  
at (c) $B=B_{c,AF}$ and (d) $B=B_{c,FP}$.}
\label{MeC}
\end{center}
\end{figure}

\begin{figure}
\begin{center}
\includegraphics*[width=0.45\textwidth,clip]{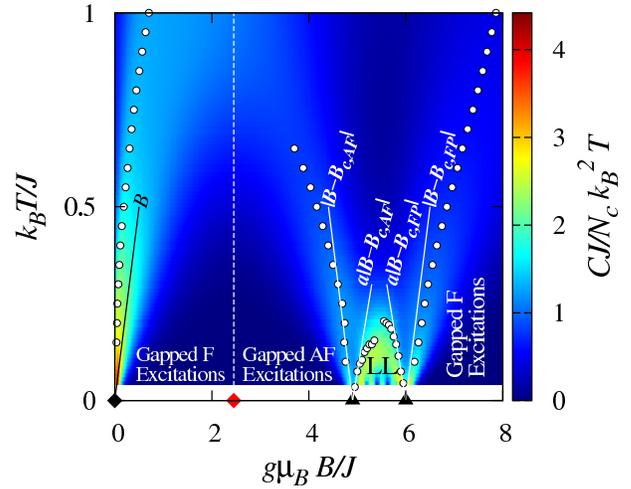}
\caption{(Color online) FTLM results for the low-$T$ phase diagram of the sS chain: 
contour plot indicates $C/T$. The critical points and crossover lines are indicated as in Fig. \ref{DiagM}.}
\label{DiagC}
\end{center}
\end{figure}
    
In Fig. \ref{DiagM} we present the contour plot of $m_c$ 
in the $T-B$ plane and a schematic phase diagram. The $T-B$ crossover lines enclosing 
the region of the LL phase, limited at $T=0$ by $B=B_{c,AF}$ and $B=B_{c,FP}$, are obtained \cite{oshikawaprl2007} from the
local extrema of $m_c(T)$ vs. $T$ for a given $B$, as shown in Fig. \ref{MeC}(a). Further, as $B\rightarrow B_c$ these crossover 
lines follow a universal function \cite{oshikawaprl2007}: $a|B-B_{c}|$ with $a=0.76238$; as shown in Fig. \ref{DiagM}, our 
numerical data confirm this asymptotic behavior for the two quantum critical points at $B=B_{c,AF}$ and $B=B_{c,FP}$.
Moreover, as $T$ increases beyond the crossover lines of the two plateaus, gapless phases are reached 
\cite{ChitraGiamarchi,sachdevprb199}. In addition, by increasing $B$ under a fixed $T$, local maxima are observed in 
the specific heat $C(B)$ per spin, as displayed in Fig. \ref{MeC}(b). These features are used to estimate \cite{giaprl2008}   
the crossover lines related to the LM plateau and FP plateau shown in Fig. \ref{DiagM}; 
in particular, we notice that $T\sim |B-B_c|$ as the lines reach the corresponding quantum
critical points \cite{ChitraGiamarchi,sachdevprb199}. 
Last, we stress that the crossover lines and the LL instability lines meet at the quantum critical points,
thus delimiting the respective quantum critical region \cite{sachdevprb199,ChitraGiamarchi,Chaboussant,oshikawaprl2007}; 
in each region  
the system is thus governed by the quantum critical point with dynamical exponent $z=2$ associated with the excited magnons, 
as discussed above. On the other hand, 
the magnon densities \cite{ChitraGiamarchi}, $n$, given by $(m_c/g\mu_B)-2$ and $3-(m_c/g\mu_B)$
for the quantum critical point at $B=B_{c,AF}$ and $B=B_{c,FP}$,
respectively, follow a universal function of $T$ and $|B-B_c|/T$: 
\begin{equation}
n=\sqrt{\frac{k_B T}{J}}f\left(\frac{|B-B_c|}{T}\right),
\end{equation}
as shown in Figs. \ref{MeC} (c) and \ref{MeC} (d). 
A better scaling behavior is observed for $B<B_{c,AF}$ ($B>B_{c,FP}$) in Fig. \ref{MeC} (c) [\ref{MeC} (d)]
since for $B>B_{c,AF}$ ($B<B_{c,FP}$) the zone of influence of the quantum critical point at $B=B_{c,FP}$ ($B=B_{c,AF}$) merges with 
the zone of influence of the point at $B=B_{c,AF}$ ($B=B_{c,FP}$). The guideline $k_B T=g\mu_B B$ in Fig. \ref{DiagM} is discussed below.

In Fig. \ref{DiagC} we present the contour plot of $C/T$ in the $T-B$ phase diagram \cite{giaprl2008}, including
the above-discussed crossover lines. At the plateaus, $C/T\rightarrow 0$
as $T\rightarrow 0$ due to the gaps, as evidenced in the plot. As we can see, the guideline $k_B T=g\mu_B B$ do 
not coincide with the local maxima of $C(B)$ in the low-$B$ 
region [see Fig. \ref{MeC}(b)] due to the LRO ferrimagnetic state at $T=0=B$: since $C\sim \sqrt{T}$,  
$C/T\rightarrow \infty$ 
as $T\rightarrow 0$ at $B=0$ and an enhancement in the intensity of $C/T$ is observed near $T=0=B$.
In spite of this fact, the plot shows a depression in the values of $C/T$ near the $T=0$ LM plateau 
which, by increasing $T$, varies in a symmetrical fashion with respect to $B=B_{m}$ (dome-shaped) and is limited 
by the $k_B T=g\mu_B B$ and $k_B T=g\mu_B |B_{c,AF}-B|$ asymptotic crossover lines. Further, the LL dome is also 
clearly seen and the crossover lines of the FP and LM gapped phases can be visualized. 
          
\begin{figure}
\begin{center}
\includegraphics[width=0.3\textwidth,clip]{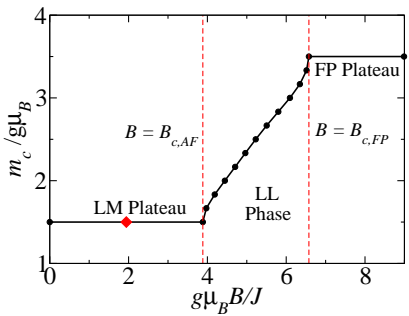}
\caption{(Color online) FTLM results for the magnetization per cell $m_c$ of the ssS chain as a function of field $B$ at $T=0$ for 
$N=18$. Full circles indicate the midpoints in the steps of the magnetization of the finite-size system and 
edges of thermodynamic-limit plateaus \cite{bonner}.}
\label{mh2}
\end{center}
\end{figure}

Next, we exhibit in Fig. \ref{mh2} the magnetization of the ssS
chain at $T=0$. For this chain, the first plateau is found at 
$m_{c}=g \mu_{B}(S-2s)$, i. e., the LM plateau, and the second is the FP plateau at $m_c=g \mu_{B}(S+2s)$; the
 LL phase is expected to occur between 
these two plateaus. A third plateau 
could be found \cite{Oshikawa} at $m_c=g\mu_B S$; however our numerical shows no evidence of this plateau. 
We remark that we did not perform a detailed analysis of the $T - B$ phase diagram
of this chain, but we expect that it should display similar features already reported for the sS chain. 

The huge values of the quantum critical magnetic fields of the CuMnDTO (sS chain) and MnNN (ssS chain) compounds, 
make the experimental investigation of the full $T-B$ phase diagram of these systems very difficult. However, 
magnetic phase transitions induced by very large magnetic fields (up to 400 T) in the low-temperature regime
have been reported \cite{highfield}. Further, materials physically described by similar models may have 
lower values for the exchange coupling and thus a more experimentally accessible phase diagram.

We also mention that ferrimagnetism can be destabilized by competing (or frustrating) interactions \cite{ivanovrev,IvanovPRB}, 
which can give rise to other critical points. Unconventional ferrimagnetism (non-bipartite lattices) 
was indeed found in one-dimensional frustrated structures \cite{AB2Frust,shimokawajpsj1} and in
the Kagom\'e lattice \cite{Hida2009}. Further, the magnetocaloric effect in the kinetically frustrated diamond 
chain was recently investigated \cite{Lyra2009}.

\section{Low-temperature magnetic susceptibility}
\begin{figure}
\includegraphics[width=0.4\textwidth,clip]{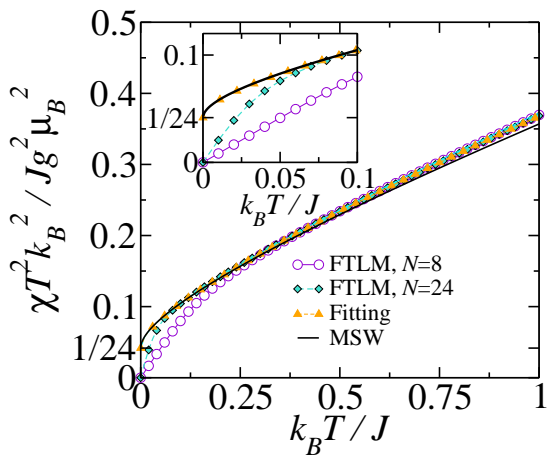}
\caption{(Color online) $\chi T^2$ vs. $T$ for the spin-1/2 ferromagnetic chain in the low and very low ($k_B T<<J$) temperature regions. 
FTLM results for $N=8$ and $N=24$.
Fitting of the FTLM results for
$N=8$ using $[\frac{1}{24}+a_0(\frac{k_B T}{J})^\frac{1}{2}+a_1(\frac{k_B T}{J})]$ (see text).
MSW results up to second order in $k_B T/J$. 
The inset shows the results for $k_B T \leq 0.1J$.}
\label{chitlinear}
\end{figure}

We now consider the temperature regime where ferromagnetic excitations tend to be a predominant feature.
In order to test and illustrate the accuracy of the FTLM in describing the susceptibility behavior at very low 
temperatures, we have calculated the susceptibility of the spin-1/2 linear \textit{ferromagnetic} chain;
the results for $\chi T^2$ as a function of $T$ are shown in Fig. \ref{chitlinear} for systems 
with 8 and 24 sites. The crossover to zero of the FTLM results as $T\rightarrow 0$ is due to finite size effects. 
We note that for $(k_B T/J) \gtrsim 0.3$ the curves for the two chain sizes superimpose,
thus suggesting that the thermodynamic-limit
behavior has already been within numerical accuracy. Also, in the temperature range $0.06\lesssim(k_B T/J) \approx 0.1$, the
 results for the larger
 system is in good agreement with the expansion formula from Takahashi's MSW theory \cite{Takahashi1987},
which up to second order in $t\equiv k_B T/J$ reads:
$\frac{\chi J}{(g\mu_B)^2}=t^{-2} \left[\frac{2}{3}s^4-2^{\frac{1}{2}}s^{\frac{5}{2}}At^\frac{1}{2}
+sA^2 t+O\left(t^{\frac{3}{2}}\right)\right]$, 
where $A=\zeta(\frac{1}{2})/\sqrt{2\pi}\approx -0.582597$ and $g=2$.
For $s=1/2$ we obtain
\begin{equation}
\frac{\chi J}{(g\mu_B)^2} = t^{-2} \left[\frac{1}{24} + 0.145649t^\frac{1}{2}
     +0.16971t+O\left(t^{\frac{3}{2}}\right)\right]
\label{Takahashi2} 
\end{equation}
We stress that in the range $ 0 < (k_B T/J) < 0.1$, Eq. (\ref{Takahashi2}) is in very good agreement with predictions 
from the Bethe-ansatz approach \cite{Takahashi1971}, while the fitting of the FTLM results for $N=8$ and $0.5<(k_B T/J)<0.9$,
yields $a_0 = 0.140$ and $a_1 = 0.186$, in good agreement with the MSW coefficients.

\begin{figure}
\includegraphics[width=0.38\textwidth,clip]{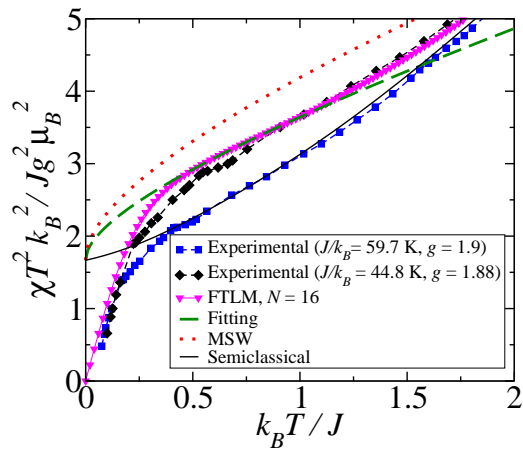}
\caption{(Color online) $\chi T^2$ vs. $T$ for the sS chain. 
Experimental data of the compound CuMnDTO from Ref. \cite{Verdaguer1984}.
FTLM results for $N=16$.
Fitting of the FTLM results for $N=16$ using $[\frac{5}{3}+a_0(\frac{k_B T}{J})^\frac{1}{2}+a_1(\frac{k_B T}{J})]$ (see text).
MSW results up to second order in $k_B T/J$ from Ref. \cite{Yamamoto1998R}. 
The semiclassical results are from Ref. \cite{Seiden1983}.}
\label{chilowtab}
\end{figure}

We now turn our attention to the low-temperature regime of the sS-chain susceptibility displayed in Fig. \ref{chilowtab}.
Firstly, we note that for $(k_B T/J) \gtrsim 0.5$ the FTLM results for $N=14$ (not shown) and $N=16$ (8 cells) coincide,
indicating that the thermodynamic limit has been attained in this temperature range.
The experimental data normalized by $J/k_B=44.8$ K ($g=1.88$) and
$J/k_B=59.7$ K ($g=1.9$) show the expected agreement with the FTLM results and the semiclassical formula, respectively,
as already displayed in Fig. \ref{chit}(a). The MSW results comes from the
expansion formula derived by Yamamoto \textit{et al.} \cite{Yamamoto1998R}, which up to second order in $t$ reads:
  \begin{eqnarray}
   \frac{\chi J}{(g\mu_B)^2}&=&t^{-2} \left[ \frac{Ss(S-s)^2}{3}-(Ss)^{\frac{1}{2}}(S-s)^{\frac{3}{2}}A t^\frac{1}{2}\right.\nonumber\\ 
                              & &\left.+(S-s)A^2t+O\left(t^{\frac{3}{2}}\right)\right].
   \label{Yamamoto}
 \end{eqnarray}

A relevant aspect of this expansion is that for $S=2s$ we recover the Takahashi expansion for the ferromagnetic linear chain
of spin $s$, which reinforces that the ferromagnetic excitation is the relevant one at low temperatures \cite{Yamamoto1998R}. Setting $s=1/2$
and $S=5/2$ in Eq. \ref{Yamamoto}, we obtain 

\begin{equation}
 \frac{\chi J}{(g\mu_B)^2} = t^{-2} \left[\frac{5}{3} + 1.842334 t^\frac{1}{2}
      +0.678 840t+O\left(t^{\frac{3}{2}}\right)\right],
\label{Yamamoto1}
\end{equation}

The FTLM results can be fitted by a function of the form $[\frac{5}{3}+a_0(\frac{k_B T}{J})^\frac{1}{2}+a_1(\frac{k_B T}{J})]$.
Guided by our studies on the spin-1/2 ferromagnetic chain, we have chosen the interval $0.5\leq (k_B T/J) \leq 0.9$ to fix the 
values of $a_0$ and $a_1$: $a_0 = 1.28$ and $a_1 = 0.69$, which can be compared with those in Eq. (\ref{Yamamoto1}),
and implies a good agreement for the integer-power coefficient and an order-of-magnitude agreement for the half-integer power coefficient.

One should notice that the FTLM results and the experimental data crossover to zero as $(k_B T/J)\rightarrow 0$, instead 
of approaching the constant value $sS(S-s)^2/3 = 5/3$.
Here one must distinguish two effects: with respect to FTLM, this is evidently
a manifestation of finite-size effects, while for the experimental data one can attribute this to the 1D/3D  crossover
that takes place below $T=7.5$ K (see Sec. III).
In fact, in the 3D region the susceptibility behaves as $\chi \sim T^{-\gamma}$, with the critical exponent $\gamma < 2$, implying
that $\chi T^2 \rightarrow 0$ as $T\rightarrow0$.

\section{Summary and Discussion}
We have presented a thorough numerical study of the GS and thermodynamic properties of two one-dimensional models related 
to quasi-one-dimensional ferrimagnetic compounds: CuMnDTO and MnNN. In fact, the models are associated to two types of ferrimagnetic 
chains: the alternating spin-1/2 spin-5/2 chain and the spin-1/2 spin-5/2 alternating chain with a spin-1/2 pendant attached
to the spin-5/2 site. The finite temperature Lanczos method proved quite reliable, except at very low temperatures where finite-size 
effects hinder its accuracy. A particular feature of these systems is the presence of 
gapless ferromagnetic and gapped AF spin-wave (magnon) branches in zero  
field. As the magnetic field is increased, the low-energy excitation changes from ferromagnetic to AF 
and the magnetic field vs. temperature phase diagram displays characteristic crossover lines which distinguish
these systems from spin-1 Haldane chains and two-leg ladder models. In particular, for the sS chain 
we have identified the quantum critical points and the crossover lines, the Luttinger liquid phase, the ferrimagnetic (LM) and
the fully polarized plateaus. The values of the exchange coupling parameters of the compounds discussed in the text
are indeed very high. 
However, magnetic phase transitions induced by very large magnetic fields (up to 400 T) in the low-temperature regime
have been experimentally investigated\cite{highfield}.
Also, other compounds described by similar models can have lower values of the exchange parameters and a more experimentally accessible phase 
diagram. We expect that this work stimulates experimental and theoretical research with focus on the phase transitions induced by
an applied magnetic field in the low-temperature regime of the large class of quasi-one-dimensional ferrimagnetic compounds.
\section{Acknowledgments}
This work was supported by CNPq, FACEPE, CAPES, and Finep (Brazilian agencies).
\bibliography{mixedchains}

\end{document}